\documentclass[12pt,a4j]{article}
\usepackage[dvips]{graphicx}
\usepackage{enumerate}
\usepackage{amsmath}
\usepackage{amssymb}
\usepackage{amsfonts}

\begin{document}
\baselineskip=7mm
\centerline{\bf B\"acklund transformation and smooth multisoliton solutions }\par
\centerline{\bf for a modified Camassa-Holm equation with cubic nonlinearity}\par
\bigskip
\centerline{Yoshimasa Matsuno\footnote{{\it E-mail address}: matsuno@yamaguchi-u.ac.jp}}\par

\centerline{\it Division of Applied Mathematical Science,}\par
\centerline{\it Graduate School of Science and Engineering} \par
\centerline{\it Yamaguchi University, Ube, Yamaguchi 755-8611, Japan} \par
\bigskip

\leftline{\bf ABSTRACT}\par
We present a compact parametric representation of the smooth bright multisolution solutions for the modified Camassa-Holm (mCH) equation
with cubic nonlinearity. We first transform the mCH equation to an associated mCH equation through a reciprocal transformation and then find a novel
B\"acklund transformation between  solutions of the associated mCH equation and a model equation for  shallow-water waves (SWW) introduced by
Ablowitz {\it at al}. We combine this result with the expressions of the multisoliton solutions for the SWW  and  modified Korteweg-de Vries equations 
 to obtain the multisoliton solutions of the mCH equation. Subsequently, we investigate the properties of the one- and two-soliton solutions as well as the
 general multisoliton solutions. We show that
 the smoothness of the solutions is assured only if the amplitude parameters of solitons satisfy certain conditions. 
 We also find that at a critical value of the parameter beyond which the solution becomes singular, the soliton solution 
 exhibits a different feature from that of the peakon solution of the CH equation.
 Then, by performing an asymptotic analysis for large time, 
 we obtain the formula for the phase shift and  confirm the solitonic nature of 
 the multisoliton solutions. Last, we use the B\"acklund transformation to derive an infinite number of  conservation laws of the mCH equation. 

\newpage

\leftline{\bf I. INTRODUCTION} \par
A modified Camassa-Holm (mCH) equation with cubic nonlinearity is a current interest in the theory of solitons. It may be written in the form
$$ m_t+[m(u^2-u_x^2)]_x=0,\quad  m=u-u_{xx}, \eqno(1.1)$$
where $u=u(x,t)$ is a real-valued function of  time $t$ and a spatial variable $x$, and the subscripts $x$ and $t$ appended to $m$ and $u$ denote partial differentiation.
The mCH equation was found independently by several researchers using a procedure that generates new integrable systems 
from  known integrable bi-hamiltonian systems.$^{1-3}$. 
It is well-known that the method yields the CH equation$^4$ when applied to the
Korteweg-de Vries (KdV) equation whereas it yields Eq. (1.1) when applied to the modified KdV (mKdV) equation.$^{1-3}$
It also arises from the physical system like nonlinear  water waves.$^{1,5}$
Although a large number of works have been devoted to  the CH equation,$^6$ the study of the
mCH equation is still  under way. 
Specifically, a recent work by Qiao reveals that it admits W/M-shape peaked solitons
unlike the peakon solutions of the CH equation.$^{5,7}$
Quite recently, smooth
 dark soliton solutions of the integrable hierarchy including the mCH equation were obtained for the variable $m$ using the inverse scattering transform (IST) method and
 the properties of the one- and two-soliton solutions were explored.$^8$ \par
The purpose of the present paper is to construct the bright $N$-soliton solutions of the mCH equation under the boundary condition $u \rightarrow u_0$ as 
$|x| \rightarrow \infty$, where $N$ is an arbitrary positive integer and $u_0$ is a positive constant. We employ an exact method of solution which worked 
effectively for constructing the $N$-soliton solutions of the CH and Degaperis-Procesi equations.$^{9-11}$ 
As a byproduct, an infinite number of conservation laws are produced in a simple manner without recourse to the IST. \par
The paper is organized as follows.    
In Sec. II, the mCH is transformed to an associated mCH equation by introducing a 
reciprocal transformation. We find a remarkable
B\"acklund transformation connecting solutions of the associated CH equation with those of  a model equation for shallow-water waves (SWW equation for short)
introduced by Ablowitz {\it et al}.$^{12}$ 
This allows us to obtain the parametric representation for the bright $N$-soliton solution of the mCH equation when coupled with the $N$-soliton
solutions of the SWW and  mKdV equations. In Sec. III, the properties of the one- and two
soliton solutions are investigated in detail.  In particular, we obtain a condition necessary for the existence of smooth solitons and
show that there exists a critical value of the amplitude parameter of the soliton beyond which the solution becomes singular.
Subsequently, the asymptotic behavior of the $N$-soliton solution is
briefly discussed, whereby we derive the formula for the phase shift and confirm the solitonic behavior of the solution.
 In Sec. IV, an infinite number of conservation laws are derived simply through the B\"acklund transformation and their explicit expressions are presented
 for the first few members.
Section V is devoted to concluding remarks. \par
\bigskip
\leftline{\bf II. CONSTRUCTION OF MULTISOLITON SOLUTIONS} \par
\leftline{\bf A. Associated mCH equation}\par
\medskip
We first introduce a coordinate transformation $(x, t)\rightarrow (y, \tau)$ by
$$dy=m\,dx-m(u^2-u_x^2)dt,\quad d\tau=dt, \eqno(2.1a)$$
subjected to the restriction $m>0$.
Accordingly, the $x$ and $t$ derivatives can be rewritten as $${\partial\over\partial x}
=m\,{\partial\over\partial y},\quad {\partial\over\partial t}=\,{\partial\over\partial \tau}-m(u^2-u_x^2){\partial\over\partial y}. \eqno(2.1b)$$
It then follows from (2.1b) that the variable $x=x(y,\tau)$ satisfies a  system of linear partial differential equations (PDEs)
$$x_y={1\over m},\quad x_\tau=u^2-m^2u_y^2. \eqno(2.2)$$
We apply the transformation (2.1b) to the mCH equation and find that it can be recast into the form
$$m_\tau+2m^3u_y=0. \eqno(2.3)$$
The integrability condition $x_{\tau y}=x_{y\tau}$ of the system (2.2) is satisfied automatically by virtue of Eq. (2.3). Using (2.3) and the relation $u_{xx}=m^2u_{yy}+mm_yu_y$,
the variable $u(=m+u_{xx})$ is expressed in terms of $m$ as
$$u=m+{1\over 2}m\left({1\over m}\right)_{\tau y}. \eqno(2.4)$$
 Last, if we define the new variable $p$ by $p=1/m$ and substitute $u$ from (2.4) into (2.3), we obtain a nonlinear PDE for $p$:
$$pp_{\tau yy}-p_yp_{\tau y}-p^3p_\tau-2p_y=0. \eqno(2.5)$$
This evolution equation for $p$ in the independent 
variables $\tau$ and $y$ is the reciprocal transformation for the mCH equation.
We call it the associated mCH equation. \par
\bigskip
\leftline{\bf B. Multisoliton solutions}\par
\medskip
The main result in this paper is given by the following theorem: \par
{\bf Theorem 2.1:} {\it  The mCH equation (1.1) admits the parametric representation
for the $N$-soliton solution 
$$u=u(y,\tau)=u_0-({\rm ln}\,f^\prime f)_{\tau y}, \eqno(2.6)$$
$$x=x(y,\tau)={y\over u_0}+u_0^2\tau+2\,{\rm ln}{f^\prime\over f}+d, \eqno(2.7)$$
where $f$ and $f^\prime$ are tau-functions given by
$$f=\sum_{\mu=0,1}{\rm exp}\left[\sum_{i=1}^N\mu_i\xi_i+\sum_{1\leq i<j\leq N}\mu_i\mu_j\gamma_{ij}\right], \eqno(2.8)$$
$$f^\prime=\sum_{\mu=0,1}{\rm exp}\left[\sum_{i=1}^N\mu_i(\xi_i-\phi_i)+\sum_{1\leq i<j\leq N}\mu_i\mu_j\gamma_{ij}\right], \eqno(2.9)$$
with
$$\xi_i=k_i\left[y-{2u_0^3\over 1-(u_0k_i)^2}\tau-\left({6\over u_0^2}-4k_i^2\right)s-y_{i0}\right],\quad (i=1, 2, ..., N),\eqno(2.10a)$$
$${\rm e}^{-\phi_i}={1-u_0k_i\over 1+u_0k_i}, \quad (i=1, 2, ..., N),\eqno(2.10b)$$
$${\rm e}^{\gamma_{ij}}=\left({k_i-k_j\over k_i+k_j}\right)^2, \quad (i, j=1, 2, ..., N; i\not=j).\eqno(2.10c)$$
\par
Here, $k_i$ and $y_{i0}$ are the amplitude and phase parameters of the $i$th soliton, respectively, $d$ is an arbitrary constant, $s$ is an auxiliary time  variable
and the notation $\sum_{\mu=0, 1}$ implies the summation over all possible combination of $\mu_1=0, 1, \mu_2=0, 1, ..., \mu_N=0, 1$. }\par
\bigskip
The parametric solution provided by Theorem 2.1 would give rise to the smooth bright $N$-soliton solution on a constant background $u=u_0$ 
if the conditions $0<u_0k_i<\sqrt{3}/2\, (i=1, 2, ..., N)$
are imposed on the amplitude parameters.
The variable $s$ introduced in (2.10a) plays the role of the time variable for both the KdV and mKdV equations. It can be set to zero after all the computations have been completed. \par
Now,  Theorem 2.1 follows straightforwardly from Propositions 2.1 and 2.2 below when combined with  the system (2.2), which we shall now demonstrate. \par
\bigskip
{\it Proposition 2.1: The variables $p, q$ and $r$ defined by
$$p={1\over u_0}+2\left({\rm ln}\,{f^\prime\over f}\right)_y, \eqno(2.11)$$
$$q=-2\,({\rm ln}\,{f})_{yy}, \eqno(2.12)$$
$$r={1\over u_0^2}-8\,({\rm ln}\,{f})_{yy}, \eqno(2.13)$$
satisfy the nonlinear PDEs
$$p_s+6p^2p_y-4p_{yyy}=0, \eqno(2.14)$$
$$q_\tau+2u_0^3\,q_y+4u_0^2\,qq_\tau+2u_0^2\,q_y\partial_y^{-1}q_\tau-u_0^2\,q_{\tau yy}=0,\quad (\partial_y^{-1}=-\int_y^\infty dy), \eqno(2.15)$$
$$r_s+6rr_y-4r_{yyy}=0, \eqno(2.16)$$
respectively. Then, the variables $r$ and $q$ are expressed in terms of $p$ as
$$r=p^2+2p_y, \eqno(2.17)$$
$$q={p_y\over 2}+{p^2\over 4}-{1\over 4u_0^2}. \eqno(2.18)$$}
\par
{\it Proof:} Equation (2.14) is the  mKdV equation of defocusing type.
 It exhibits the dark $N$-soliton solution (2.11) with the tau-functions
 (2.8) and (2.9).$^{13,14}$
Equation (2.15) is a SWW equation for which the $N$-soliton solution is given by (2.12) with the tau-function (2.8) whereas Eq. (2.16) is the KdV equation
whose  tau-function has the same functional form as that of the SWW equation except the time dependence.$^{12,15}$ 
The latter follows from the fact that the SWW equation belongs to a member
of the integrable hierarchy of the KdV equation.$^{12}$ \par
The relation (2.17) is  the Miura transformation$^{16}$ which connects solutions of the
mKdV equation with those of the KdV equation.
Actually,  a direct substitution of (2.17) into the left-hand side of (2.16) reveals that
$$r_s+6rr_y-4r_{yyy}=2\left({\partial\over\partial y}+p\right)(p_s+6p^2p_y-4p_{yyy}).\eqno(2.19)$$
\par
The relation (2.18) stems simply from (2.12), (2.13) and (2.17) by eliminating the tau-function  $f$. \hspace{\fill}$\Box$ \par
\medskip
The relation (2.18) is a B\"acklund transformation between solutions of the associated mCH equation and the SWW equation, as
will be shown by Proposition 2.2. 
If one substitutes (2.11)-(2.13) into (2.17) and (2.18), then one can see that they reduce to the following bilinear equation for $f$ and $f^\prime$:
$$f_{yy}^\prime f-2f_y^\prime f_y +f^\prime f_{yy}+{1\over u_0}(f_y^\prime f-f^\prime f_y)=0. \eqno(2.20)$$
The above equation is a constituent of the system of bilinear equations for the mKdV equation
 and hence the
tau-functions $f$ and $f^\prime$ from (2.8) and (2.9) solve Eq. (2.20).$^{17}$
\par
\bigskip
{\it Proposition 2.2: Let
$$Q=q_\tau+2u_0^3\,q_y+4u_0^2\,qq_\tau+2u_0^2\,q_y\partial_y^{-1}q_\tau-u_0^2\,q_{\tau yy}, \eqno(2.21)$$
$$R={p_{\tau yy}\over p}-{p_yp_{\tau y}\over p^2}-pp_\tau-2\,{p_y\over p^2}. \eqno(2.22)$$
Then, $Q$ and $R$ are connected by the relation
$$Q=-{u_0^2\over 2}\left({\partial\over\partial y}+p\right){\partial\over\partial y}(p\,\partial_y^{-1}R), \eqno(2.23)$$ 
under the transformation (2.18).}
\par
\bigskip
{\it Proof:} Substituting (2.18) into (2.21), one can recast $Q$ into the form
$$Q={u_0^2\over 2}[(p_{yy}+pp_y)\{2u_0+\partial_y^{-1}(pp_\tau)\}+p^2p_{\tau y}+3pp_\tau p_y+p^3p_\tau-p_{\tau yyy}-pp_{\tau yy}].$$
On the other hand, integration of (2.22) with respect to $y$ under the boundary condition $p\rightarrow 1/u_0, \ |y|\rightarrow\infty$ gives
$$2u_0+\partial_y^{-1}(pp_\tau)=-\partial_y^{-1}R+{1\over p}(p_{\tau y}+2),$$
whereas a direct computation using (2.22) leads to
$${(p^2R)_y\over p}+p^2R=-p^2p_{\tau y}-3pp_\tau p_y-p^3p_\tau+p_{\tau yyy}+pp_{\tau yy} -{1\over p}(p_{yy}+pp_y)(p_{\tau y}+2).$$
The relation (2.22) follows by introducing above two expressions into $Q$. Actually,
$$Q=-{u_0^2\over 2}[(p_{yy}+pp_y)\partial_y^{-1}R+pR_y+(2p_y+p^2)R]
=-{u_0^2\over 2}\left({\partial\over\partial y}+p\right){\partial\over\partial y}(p\,\partial_y^{-1}R).$$
This completes the proof. \hspace{\fill}$\Box$ \par
\medskip
It is obvious from (2.23) that the equation $R=0$ yields the equation $Q=0$. But, the converse statement is not true in general.
For the $N$-soliton solution (2.11) which satisfies the boundary condition $p\rightarrow p_0\,(=1/u_0), |y|\rightarrow \infty$, however, $p$ is a solution
of Eq. (2.5), or $R=0$ if $q$ from (2.12) satisfies Eq. (2.15), or $Q=0$.
To see this, assume first that $Q=0$. Integrating the resultant equation from (2.23) with respect to $y$ and using the 
boundary condition $p\rightarrow p_0, y\rightarrow +\infty$, one obtains
$${\partial\over\partial y}(p\,\partial_y^{-1}R)=R_0(\tau)\,{\rm exp}\left[\int_y^\infty(p-p_0)dy-p_0y\right], \eqno(2.24)$$
where $R_0(\tau)$ is an integration constant.
A slight modification of the relation for $\partial_y^{-1}R$ derived in the proof of Proposition 2.2  gives
$$\partial_y^{-1}R={p_{\tau y}\over p}+{1\over 2}{\partial\over\partial\tau}\int_y^\infty(p^2-p_0^2)dy+{2\over p}-{2\over p_0}.$$
As confirmed easily by integrating (2.18) with respect to $y$ and using (2.12), the integral $\int_{-\infty}^\infty(p^2-p_0^2)dy$ becomes a constant independent of $\tau$.
It turns out that $\lim_{y\rightarrow -\infty}\partial_y^{-1}R=0$. With this fact in mind,
 the left-hand side of (2.24) vanishes in the limit of $y\rightarrow -\infty$. On the other hand, the right-hand side diverges in this limit.
Thus, one must put $R_0=0$, which yields $(p\,\partial_y^{-1}R)_y=0$.
Integrating again with respect to $y$, one has $\partial_y^{-1}R=R_1(\tau)/p$, where $R_1(\tau)$ is an integration constant.
In view of the boundary condition, this constant must be zero, and hence $\partial_y^{-1}R=0$. Differentiating this relation by $y$,
one finally arrives at the relation $R=0$. \par
\bigskip
We are now ready for  proving Theorem 1.1. \par
{\it Proof:}\ First, referring to (2.2) and (2.11), one obtains
$$x_y=p={1\over u_0}+2\left({\rm ln}{f^\prime\over f}\right)_y.$$
Integrating this relation with respect to $y$ leads to
$$x={y\over u_0}+{\rm ln}{f^\prime\over f}+d(\tau),$$
where $d(\tau)$ is an integration constant. Differentiate the  above expression by $\tau$ and use (2.2) to find the relation
$$x_\tau=2\left({\rm ln}{f^\prime\over f}\right)_\tau+d^\prime(\tau)=u^2-m^2u_y^2.$$
This expression reduces, upon taking the limit $y\rightarrow \infty$ and substituting the limiting values  $u\rightarrow u_0, u_y\rightarrow 0, [{\rm ln}(f^\prime/f)]_\tau\rightarrow 0$,
to $d^\prime(\tau)=u_0^2$, which
gives $d(\tau)=u_0^2\,\tau+d$ after integrating with respect to $\tau$. This proves (2.7). \par
 It now follows by integrating Eq. (2.3) with respect to $y$ and substituting $p^2$ from (2.18)
into the resultant expression that
$$u=u_0+{1\over 4}\int_{-\infty}^y(p^2)_\tau dy=u_0+\int_{-\infty}^yq_\tau dy-{1\over 2}\,p_\tau. $$
Last, substitution of (2.11) and (2.12) into the above expression yields
$$u=u_0-2({\rm ln}\, f)_{\tau y}-\left({\rm ln}\,{f^\prime\over f}\right)_{\tau y}=u_0-({\rm ln}\,f^\prime f)_{\tau y},$$
which is just (2.6). \hspace{\fill}$\Box$ \par
\bigskip
\leftline{\bf III. PROPERTIES OF SOLUTIONS} \par
\leftline{\bf A. One-soliton solution}\par
\medskip
The tau-functions $f$ and $f^\prime$ corresponding to the one-soliton solution are given by (2.8), (2.9) and (2.10) with $N=1$.  They read
$$f=1+{\rm e}^\xi, \eqno(3.1)$$
$$f^\prime=1+{\rm e}^{\xi-\phi}, \eqno(3.2)$$
with
$$\xi=k(y-\tilde c\tau-y_0),\quad \tilde c={2u_0^3\over 1-(u_0k)^2},\eqno(3.3a)$$
$${\rm e}^{-\phi}={1-u_0k\over 1+u_0k}, \eqno(3.3b)$$
where we have put $\xi=\xi_1, \phi=\phi_1, k=k_1$ and $y_0=y_{10}$ for simplicity and set $s=0$.  We assume $k>0$  hereafter. \par
The parametric representation of the one-soliton solution follows by introducing (3.1) and (3.2) into (2.6) and (2.7).  It may be written in the form
$$u=u_0+{k^2\tilde c\{a\cosh(\xi-\xi_0)+1\}\over \{\cosh(\xi-\xi_0)+a\}^2}, \eqno(3.4a)$$
$$x-ct-x_0={\xi\over u_0k}+2\,{\rm ln}\left({\sqrt{1-\alpha}\,{\rm e}^{\xi-\xi_0}+\sqrt{1+\alpha}
\over \sqrt{1+\alpha}\,{\rm e}^{\xi-\xi_0}+\sqrt{1-\alpha}}\right), \eqno(3.4b)$$
where 
$$\alpha=u_0k,\quad a={1\over \sqrt{1-(u_0k)^2}}, \quad \xi_0={1\over 2}\,{\rm ln}\left({1+u_0k\over 1-u_0k}\right), \eqno(3.4c)$$
$$c={\tilde c\over u_0}+u_0^2={u_0^2\{3-(u_0k)^2\}\over 1-(u_0k)^2}, \eqno(3.4d)$$
and we have put $d=x_0-y_0/u_0-{\rm ln}[(1-\alpha)/(1+\alpha)]$. Note that $c$ from (3.4d) is the velocity of the soliton in the $(x, t)$ coordinate system.
The smoothness of the soliton solution is assured by  the condition $m=1/p>0$ (see (2.1a) and (2.1b)) which imposes certain restriction on the parameter $\alpha$.
To show this explicitly, we compute $x_y(=p)$ from (3.4b) and obtain
$$p={1\over u_0}-{2k\alpha\over \sqrt{1-\alpha^2}\,\cosh(\xi-\xi_0)+1}. \eqno(3.5)$$
Recall that $p$ is a dark soliton solution of the mKdV equation (2.14). 
The required condition is then found to be satisfied if $1/u_0-2k\alpha/(\sqrt{1-\alpha^2}+1)>0$. Thus, the parameter $\alpha$ must lie in the interval
$$0<\alpha<{\sqrt{3}\over 2}. \eqno(3.6)$$
\par
One can see from (3.4a) and (3.4b) that the  one-soliton solution represents a bright soliton on a constant background $u=u_0$ 
whose center position $x_c$   is located at $x_c=ct+x_0+\xi_0/(u_0k)\ (\xi=\xi_0)$.
In view of this observation,  the amplitude of
the soliton with respect to the background field, which we denote by $A$, amounts to
$$A={2u_0\over\sqrt{1-\alpha^2}}(1-\sqrt{1-\alpha^2}). \eqno(3.7)$$
Eliminating the parameter $\alpha$ from (3.4d) and (3.7) leads to the amplitude-velocity relation of the soliton
$$A=\sqrt{2(c-u_0^2)}-2u_0.\eqno(3.8)$$
The inequality (3.6) restricts  possible values of $c$ and $A$. Explicitly,
$$3u_0^2<c<9u_0^2,\quad 0<A<2u_0. \eqno(3.9)$$
\par
\begin{figure}[t]
\begin{center}
\includegraphics[width=10cm]{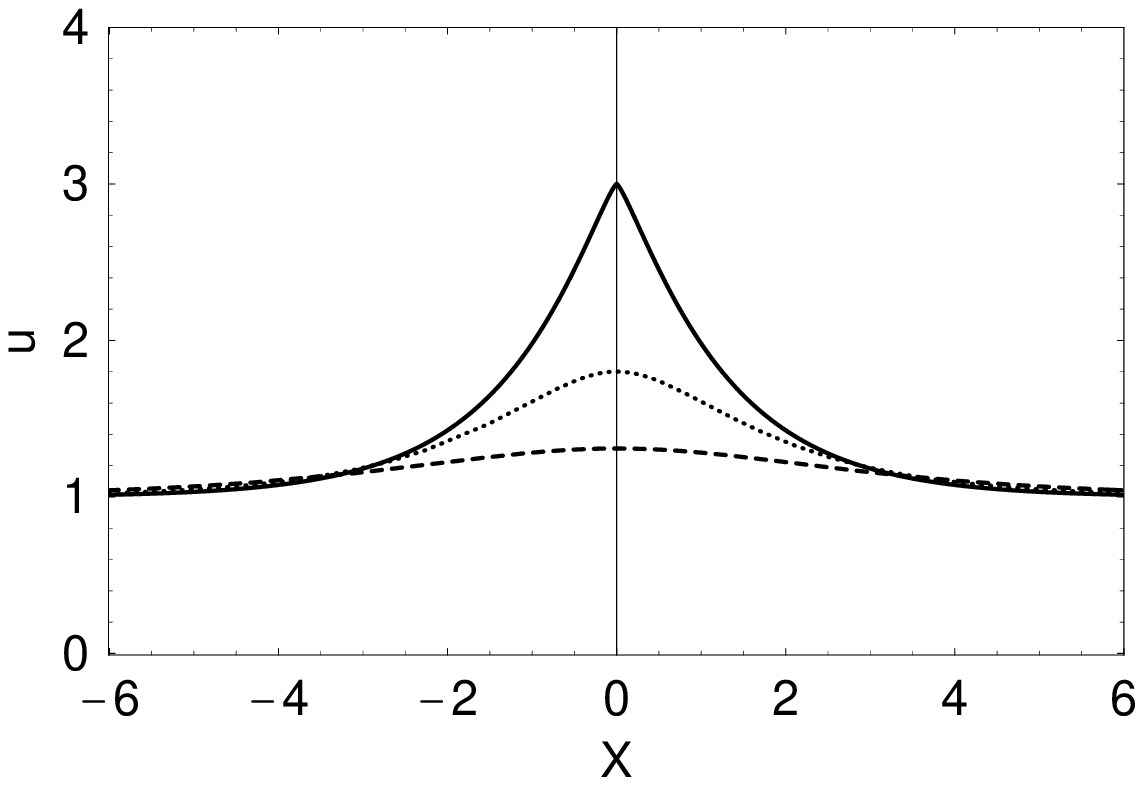}
\end{center}
\noindent {{\bf FIG. 1.}\ The profile of  smooth solitons. $\alpha=0.5$\,(dashed curve), $\alpha=0.7$\,(dotted curve), $\alpha=\sqrt{3}/2$\,(solid curve).}
\end{figure}

If the condition (3.6) is satisfied, then the one-soliton solution represents a smooth bright solitary wave travelling to the right with
 velocity $c$ and amplitude $A$. 
 As $\alpha$ increases, the amplitude grows and the width narrows. This observation is illustrated in Fig. 1, whereby
 the profile of smooth solitons with a background $u_0=1$  is depicted 
agianst the stationary variable $X=x-ct-x_0$ for three representative values of $\alpha$. 
Of particular interest is the limiting profile of the soliton when the parameter $\alpha$ tends to the upper limit $\alpha=\sqrt{3}/2\,(\simeq 0.866)$ 
of the inequality (3.6) \ (see solid curve in Fig. 1).
In this limit, the smoothness of the profile would be lost at the center position (or the crest) of the soliton. 
The typical profile is also depicted in Fig. 2 for  singular solitons which violate the condition (3.6).
\par
To explore the feature of the solution in more detail, we shall investigate the profile of the soliton near its crest. 
To this end, we first derive the approximate expression of $X$  from (3.4b).
If we choose the phase parameter of the soliton appropriately such that $\xi=\xi_0$ corresponds to $X=0$, then
the  power series expansion  of $X$ with respect to $\xi-\xi_0$ is found to be as
$$X=-{1\over \alpha}(1-2\,\sqrt{1-\alpha^2})\,(\xi-\xi_0)+{1\over 3}{\sqrt{1-\alpha^2}\over \alpha^3}\left(1-\sqrt{1-\alpha^2}\right)^2(\xi-\xi_0)^3+O\left((\xi-\xi_0)^5\right). \eqno(3.10)$$

\begin{figure}[t]
\begin{center}
\includegraphics[width=10cm]{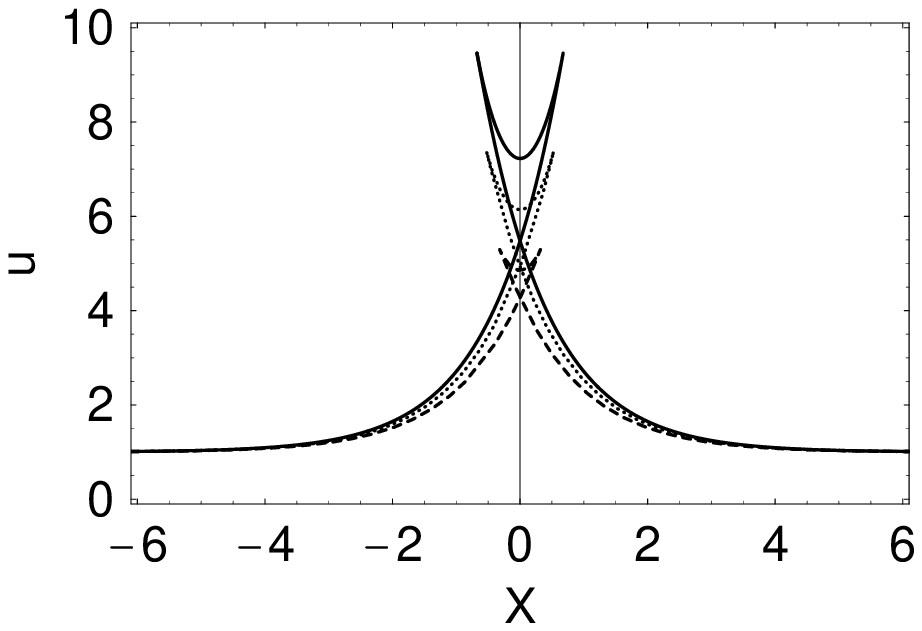}
\end{center}
\noindent{{\bf FIG. 2.}\ The profile of  singular solitons. $\alpha=0.94$\,(dashed curve), $\alpha=0.96$\,(dotted curve), $\alpha=0.97$\,(solid curve).}

\end{figure}

\noindent In particular, for $\alpha=\sqrt{3}/2$, (3.10) reduces to
$$X= {\sqrt{3}\over 27}\,(\xi-\xi_0)^3+O\left((\xi-\xi_0)^5\right). \eqno(3.11) $$
\par
On the other hand, the expansion of $u$ from (3.4a) reads
$$u=u_0\left[1+{6\over 1+a}+{3(-2+a)\over (1+a)^2}\,(\xi-\xi_0)^2+{16-13a+a^2\over 4(1+a)^3}\,(\xi-\xi_0)^4+O\left((\xi-\xi_0)^6\right)\right], \eqno(3.12)$$
which simplifies to
$$u=u_0\left[3-{1\over 18}(\xi-\xi_0)^4+O\left((\xi-\xi_0)^6\right)\right]. \eqno(3.13)$$
when $a=2\,(\alpha=\sqrt{3}/2)$.
Note that the coefficient of $(\xi-\xi_0)^2$ in the expansion (3.12) vanishes for this specific value of $a$.
By inserting $\xi-\xi_0$ from (3.11) into (3.13), we  find  that the profile of the soliton near the crest is approximated by the
expression
$$u \sim u_0\left[3-{3^{4/3}\over 2}\,\,X^{4/3}\right]. \eqno(3.14)$$
\par
Thus, we can see that the smoothness of the solution is lost at the crest $X=0$. More precisely,  $u$ has only a continuous first derivative near the crest 
and  the $n$th derivative does not exists for $n\geq 2$ at the crest.
This novel feature is in striking contrast to the appearance of the singularity (or discontinuous first derivative at the crest) in the peakon solution which can be deduced from
 the smooth soliton solution of the CH equation through an appropriate limiting process.$^{18,19}$
\par
\bigskip
\leftline{\bf B. Two-soliton solution}\par
\medskip
The tau-functions $f$ and $f^\prime$ for the two-soliton solution can be written as
$$f=1+{\rm e}^{\xi_1}+{\rm e}^{\xi_2}+\left({k_1-k_2\over k_1+k_2}\right)^2{\rm e}^{\xi_1+\xi_2},\eqno(3.15a)$$
$$f^\prime=1+{\rm e}^{\xi_1-\phi_1}+{\rm e}^{\xi_2-\phi_2}+\left({k_1-k_2\over k_1+k_2}\right)^2{\rm e}^{\xi_1+\xi_2-\phi_1-\phi_2},\eqno(3.15b)$$
where
$$\xi_i=k_i(y-\tilde c_i\tau-y_{i0}),\quad \tilde c_i={2u_0^3\over 1-(u_0k_i)^2},\quad (i=1, 2),\eqno(3.15c)$$
$${\rm e}^{-\phi_i}={1-u_0k_i\over 1+u_0k_i},\quad (i=1, 2). \eqno(3.15d)$$
We investigate the asymptotic behavior of the two-soliton solution for large time and show that it describes the elastic interaction of two
bright solitons on a background field $u=u_0$. To this end, let $c_i\ (i=1, 2)$ be the velocity of the $i$the soliton in the $(x, t)$ coordinate system
and assume that $0<c_2<c_1$. 
In addition, we impose the conditions $0<u_0k_i<\sqrt{3}/2,\ (i=1, 2)$ to assure the smoothness of the solution.
\par
First, we take the limit  $t\rightarrow -\infty$ with $\xi_1$ being fixed. In this limit, $\xi_2\rightarrow-\infty$. 
Then, the leading-order asymptotics of the tau-functions are given by
$$f \sim 1+{\rm e}^{\xi_1}, \eqno(3.16a)$$
$$f^\prime=1+{\rm e}^{\xi_1-\phi_1}. \eqno(3.16b)$$
It folows from (2.6) and (2.7), (3.16a), and (3.16b) that
$$u \sim u_0+{k_1^2\tilde c_1\{a_1\cosh(\xi_1-\xi_{10})+1\}\over \{\cosh(\xi_1-\xi_{10})+a_1\}^2}, \eqno(3.17a)$$
$$x-c_1t-x_{10} \sim{\xi_1\over u_0k_1}+2\,{\rm ln}\left({\sqrt{1-\alpha_1}\,{\rm e}^{\xi_1-\xi_{10}}+\sqrt{1+\alpha_1}
\over \sqrt{1+\alpha_1}\,{\rm e}^{\xi_1-\xi_{10}}+\sqrt{1-\alpha_1}}\right), \eqno(3.17b)$$
where 
$$\alpha_1=u_0k_1,\quad a_1={1\over \sqrt{1-(u_0k_1)^2}}, \quad \xi_{10}={1\over 2}\,{\rm ln}\left({1+u_0k_1\over 1-u_0k_1}\right), \eqno(3.17c)$$
$$c_1={\tilde c_1\over u_0}+u_0^2={u_0^2\{3-(u_0k_1)^2\}\over 1-(u_0k_1)^2}, \eqno(3.17d)$$
\par
In the limit $t\rightarrow +\infty$,  on the other hand, $\xi_2\rightarrow +\infty$. The expressions corresponding to (3.16a)-(3.17b) read
$$f \sim {\rm e}^{\xi_2}\left[1+\left({k_1-k_2\over k_1+k_2}\right)^2{\rm e}^{\xi_1}\right], \eqno(3.18a)$$
$$f^\prime={\rm e}^{\xi_2-\phi_2}\left[1+\left({k_1-k_2\over k_1+k_2}\right)^2{\rm e}^{\xi_1-\phi_1}\right]. \eqno(3.18b)$$
$$u \sim u_0+{k_1^2\tilde c_1\{a_1\cosh(\xi_1-\xi_{10}+\delta_1^{(+)})+1\}\over \{\cosh(\xi_1-\xi_{10}+\delta_1^{(+)})+a_1\}^2}, \eqno(3.19a)$$
$$x-c_1t-x_{10} \sim {\xi_1\over u_0k_1}+2\,{\rm ln}\left({\sqrt{1-\alpha_1}\,{\rm e}^{\xi_1-\xi_{10}+\delta_1^{(+)}}+\sqrt{1+\alpha_1}
\over \sqrt{1+\alpha_1}\,{\rm e}^{\xi_1-\xi_{10}+\delta_1^{(+)}}+\sqrt{1-\alpha_1}}\right)-2\phi_2, \eqno(3.19b)$$
where
$$\delta_1^{(+)}={\rm ln}\left({k_1-k_2\over k_1+k_2}\right)^2. \eqno(3.19c)$$
\par
Let $x_{ic}$ be the center position of the $i$th soliton. Then, as $t\rightarrow -\infty$, we find
$$x_{1c}\sim c_1t+x_{10}+{\xi_{10}\over u_0k_1},\quad (\xi_1=\xi_{10}). \eqno(3.20)$$
As $t\rightarrow +\infty$, on the other hand, $x_{1c}$ reads
$$x_{1c}\sim c_1t+x_{10}+{{1\over u_0k_1}(\xi_{10}-\delta_1^{(+)})}-2\phi_2,\quad (\xi_1=\xi_{10}-\delta_1^{(+)}). \eqno(3.21)$$
\par
The above analysis shows that the asymptotic state of the solution for large time is represented by a superposition of two single solitons in the
rest frame of reference. The net effect of the interaction between solitons is the phase shift, which we shall now evaluate. To this end,
we define the phase shift of the $i$th soliton by 
$$\Delta_i=x_{ic}(t\rightarrow+\infty)-x_{ic}(t\rightarrow-\infty),\quad (i=1, 2). \eqno(3.22)$$
Then, we see from (3.20) and (3.21) that the large soliton suffers a phase shift
$$\Delta_1=-{1\over u_0k_1}{\rm ln}\left({k_1-k_2\over k_1+k_2}\right)^2-{\rm ln}\left({1+u_0k_2\over 1-u_0k_2}\right)^2. \eqno(3.23)$$
Performing the similar asymptotic analysis for the small soliton, we obtain the formula for the phase shift. We quote only the final result.
$$\Delta_2={1\over u_0k_2}{\rm ln}\left({k_1-k_2\over k_1+k_2}\right)^2+{\rm ln}\left({1+u_0k_1\over 1-u_0k_1}\right)^2. \eqno(3.24)$$
\par

\begin{figure}[t]
\begin{center}
\includegraphics[width=10cm]{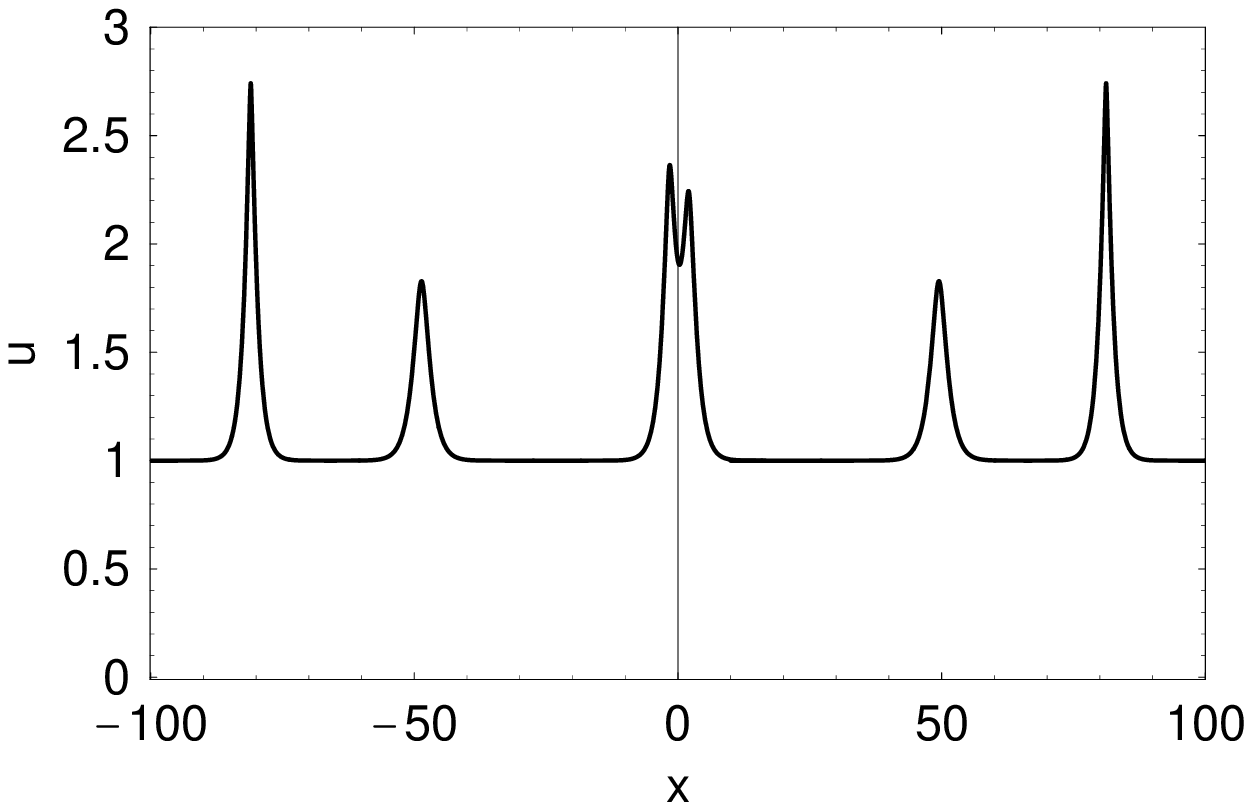}
\end{center}
\noindent{\bf FIG. 3.}\ The interaction of two solitons. $t=-10$\,(left: before collision), $t=0$\,(middle: during collision), $t=10$\,(right: after collision).\par
\end{figure}

The first term of the above formulas is the same as the phase shift arising from the interaction of two solitons for the KdV equation and the second term comes from the mapping (2.7).
A new feature appears due to this additional term. An inspection reveals that under the condition $c_2<c_1\ (k_2<k_1)$, $\Delta_1$ is always positive whereas $\Delta_2$ takes any sign
depending on values of $k_1$ and $k_2$. Specifically, there exists a critical curve along which $\Delta_1=\Delta_2$ and beyond which $\Delta_1<\Delta_2$ so that
the small soliton exhibits a larger phase shift than that of the large soliton. 
It is interesting to observe that the formulas (3.23) and (3.24) coincide formally with those of the two-soliton solution of the CH equation
if one identifies $u_0$ with the parameter $\kappa$ which enters into a linear dispersive term $2\kappa^2 u_x$ in the CH equation.$^9$ \par
 The interaction process of  two solitons  is  illustrated in Fig. 3 for three values of time.
In this example, $u_0=1, c_1=0.8, c_2=0.5, A_1=1.74, A_2=0.83$. The phase shifts of the large and small solitons evaluated from (3.23) and (3.24)
amount to  $\Delta_1=2.20$ and  $\Delta_2=-1.89$, respectively.
\par
\bigskip
\leftline{\bf C. {\it N}-soliton solution}\par
\medskip
We discuss briefly the asymptotic behavior of the general $N$-soliton solution. First, we order the magnitude of the velocity of each soliton as
$0<c_N<c_{N-1}<...<c_1$ and impose the conditions  $0<u_0k_i<\sqrt{3}/2,\ (i=1, 2, ..., N)$. 
We take the limit $t\rightarrow-\infty$ with the coordinate $\xi_i$ of the $i$th soliton being fixed. Then, $\xi_1, \xi_2, ..., \xi_{i-1}\rightarrow+\infty$,
and $\xi_{i+1}, \xi_{i+2}, ..., \xi_N\rightarrow-\infty$. The asymptotic forms of $f$ and $f^\prime$ from (2.8) and (2.9) are found to be as
$$f\sim\prod_{1\leq j<l\leq i-1}\left({k_j-k_l\over k_j+k_l}\right)^2{\rm exp}\left[\sum_{j=1}^{i-1}\xi_j\right]\left(1+{\rm e}^{\xi_i+\delta_i^{(-)}}\right),\eqno(3.25a)$$
$$f^\prime\sim\prod_{1\leq j<l\leq i-1}\left({k_j-k_l\over k_j+k_l}\right)^2{\rm exp}\left[\sum_{j=1}^{i-1}(\xi_j-\phi_j)\right]\left({\rm e}^{\xi_i-\phi_i+\delta_i^{(-)}}\right),\eqno(3.25b)$$
where
$$\delta_i^{(-)}=\sum_{j=1}^{i-1}{\rm ln}\left({k_i-k_j\over k_i+k_j}\right)^2. \eqno(3.25c)$$
Substitution of (3.25a) and (3.25b) into (2.6) and (2.7) yields
$$u \sim u_0+{k_i^2\tilde c_i\{a_i\cosh(\xi_i-\xi_{i0}+\delta_i^{(-)})+1\}\over \{\cosh(\xi_i-\xi_{i0}+\delta_i^{(-)})+a_i\}^2}, \eqno(3.26a)$$
$$x-c_it-x_{i0} \sim {\xi_i\over u_0k_i}+2\,{\rm ln}\left({\sqrt{1-\alpha_i}\,{\rm e}^{\xi_i-\xi_{i0}+\delta_i^{(-)}}+\sqrt{1+\alpha_i}
\over \sqrt{1+\alpha_i}\,{\rm e}^{\xi_i-\xi_{i0}+\delta_i^{(-)}}+\sqrt{1-\alpha_i}}\right)
-2\sum_{j=1}^{i-1}\phi_j. \eqno(3.26b)$$
\par
In the limit $t\rightarrow+\infty$, on the other hand, 
$\xi_1, \xi_2, ..., \xi_{i-1}\rightarrow-\infty$,
and $\xi_{i+1}, \xi_{i+2}, ..., \xi_N\rightarrow+\infty$ and the expressions corresponding to (2.25) and (2.26) are given by
$$f\sim\prod_{i+1\leq j<l\leq N}\left({k_j-k_l\over k_j+k_l}\right)^2{\rm exp}\left[\sum_{j=i+1}^N\xi_j\right]\left(1+{\rm e}^{\xi_i+\delta_i^{(+)}}\right),\eqno(3.27a)$$
$$f^\prime\sim\prod_{i+1\leq j<l\leq N}\left({k_j-k_l\over k_j+k_l}\right)^2{\rm exp}\left[\sum_{j=i+1}^{N}(\xi_j-\phi_j)\right]\left(1+{\rm e}^{\xi_i-\phi_i+\delta_i^{(+)}}\right),\eqno(3.27b)$$
where
$$\delta_i^{(+)}=\sum_{j=i+1}^{N}{\rm ln}\left({k_i-k_j\over k_i+k_j}\right)^2, \eqno(3.27c)$$
and
$$u \sim u_0+{k_i^2\tilde c_i\{a_i\cosh(\xi_i-\xi_{i0}+\delta_i^{(+)})+1\}\over \{\cosh(\xi_i-\xi_{i0}+\delta_i^{(+)})+a_i\}^2}, \eqno(3.28a)$$
$$x-c_it-x_{i0} \sim {\xi_i\over u_0k_i}+2\,{\rm ln}\left({\sqrt{1-\alpha_i}\,{\rm e}^{\xi_i-\xi_{i0}+\delta_i^{(+)}}+\sqrt{1+\alpha_i}
\over \sqrt{1+\alpha_i}\,{\rm e}^{\xi_i-\xi_{i0}+\delta_i^{(+)}}+\sqrt{1-\alpha_i}}\right)
-2\sum_{j=i+1}^{N}\phi_j. \eqno(3.28b)$$
\par
We can see that the asymptotic form of the $N$-soliton solution is a superposition of $N$ single solitons each of which has the form given by (3.4).
The phase shift of the $i$th soliton can be derived from (3.26b) and (3.28b). It reads
$$\Delta_i={1\over u_0k_i}\left[\sum_{j=1}^{i-1}{\rm ln}\left({k_i-k_j\over k_i+k_j}\right)^2-\sum_{j=i+1}^{N}{\rm ln}\left({k_i-k_j\over k_i+k_j}\right)^2\right]$$
$$+\sum_{j=1}^{i-1}\left({1+u_0k_j\over1-u_0k_j}\right)^2-\sum_{j=i+1}^{N}\left({1+u_0k_j\over1-u_0k_j}\right)^2,\quad (i=1, 2, ..., N). \eqno(3.29)$$
In the special case of $N=2$, the formulas (3.29) reduce to (3.23) and (3.24).
They clearly show that each soliton has pairwise interactions with other
solitons, i.e., there are no many-particle collisions among solitons. This feature is common
to that of the $N$-soliton solutions of integrable nonlinear PDEs.
\par
\bigskip
\leftline{\bf IV. CONSERVATION LAWS}\par
\bigskip
The B\"acklund transformation (2.18) between solutions $p$ and $q$ of the two integrable equations allows us to construct an infinite number of conservation laws of
the mCH equation in a simple manner. To demonstrate this,  we first note that the SWW equation (2.15) exhibits local conservation laws of the form
$$w_{n,\tau}=j_{n,y},\quad (n=0, 1, 2, ..), \eqno(4.1)$$
where the conserved density $w_n$ and associated flux $j_n$ are polynomials of $q$ and its $y$-derivatives. We rewrite (4.1) in terms of the variables $x$ and $t$ by using (2.1). Substituting Eq. (1.1)
into the resultant equation, we obtain
$$(mw_n)_t=[j_n-m(u^2-u_x^2)w_n]_x. \eqno(4.2)$$
It turns out that the quantities
$$I_n=\int_{-\infty}^\infty mw_ndx,\quad 
(n=0, 1, 2, ...), \eqno(4.3)$$
become the conservation laws of the mCH equation if one introduces $q$ from (2.18) into $w_n$. \par
The conservation laws of the SWW equation take the same form as those of the KdV equation since both
equations belong to a common integrable hierarchy. For completeness, we reproduce them shortly.
We start from the Lax pair associated with the SWW equation$^{20}$
$$\psi_{yy}-q\psi={\lambda\over u_0^2}\psi, \eqno(4.4a)$$
$$(4\lambda-1)\psi_\tau=2u_0^2(u_0+\partial_y^{-1}q_\tau)\psi_y-u_0^2q_\tau\psi, \eqno(4.4b)$$
where $\lambda$ is a spectral parameter.
If we put $w=\psi_y/\psi$, we can rewrite  (4.4) into the form
$$w_y+w^2-q={\lambda\over u_0^2}, \eqno(4.5a)$$
$$(4\lambda-1)w_\tau=[2u_0^2(u_0+\partial_y^{-1}q_\tau)w-u_0^2q_\tau]_y. \eqno(4.5b)$$
Thus, the quantity $\int_{-\infty}^\infty w\,dy$ is conserved in $\tau$.
Expanding $w$ in powers of $\epsilon(\equiv u_0/(2\sqrt{\lambda}))$ as $w=1/(2\epsilon)+\sum_{n=1}^\infty\epsilon^nw_n$ 
and substituting it into (4.5a), we obtain, after equating the coefficients of $\epsilon^n$, the recursion relation
that determines $w_n$
$$w_{n+1}=w_{n,y}-\sum_{m=1}^{n-1}w_{n-m}w_m,\quad (n\geq 2), \eqno(4.6)$$
with the initial conditions $w_1=q$ and $w_2=-q_y$. We quote a few  conserved densities:
$$w_3=q_{yy}-q^2, \quad w_4=(-q_{yy}+2q^2)_y,\quad w_5=(q_{yyy}-6qq_y)_y+2q^3+q_y^2. \eqno(4.7)$$
The quantity $q$ from (2.18) can be rewritten in terms of $m$ as
$$q=-{m_y\over 2m^2}+{1\over 4m^2}-{1\over 4u_0^2}. \eqno(4.8)$$
\par
Last, substituting (4.7) with (4.8) into (4.3) and taking into account the relation $\partial/\partial y=m^{-1}\partial/\partial x$, we obtain 
a sequence of conservation laws of the mCH equation expressed in terms of the original spatial variable.
We see that $w_n$ for even $n$ takes the form of a perfect derivative $(W_n)_y$ and hence the quantity $I_n$ for even $n$ vanishes identically. Actually, 
$\int_{-\infty}^\infty m(W_n)_ydx=\int_{-\infty}^\infty (W_n)_xdx=0$.
Hence, only the quantities $I_n$ for odd $n$ survive. We can express the nontrivial conservation laws as $I_{2n+1}=\sum_{m=0}^{n+1}\nu_{nm}\tilde I_m$,
where $\nu_{nm}$ are  constants depending on $u_0$.
The first four of $\tilde I_n$ read as follows:
$$\tilde I_0=\int_{-\infty}^\infty(m-u_0)dx, \eqno(4.9a)$$
$$\tilde I_1=\int_{-\infty}^\infty\left({1\over m}-{1\over u_0}\right)dx,\eqno(4.9b)$$
$$\tilde I_2=\int_{-\infty}^\infty\left[{1\over m^3}-{1\over u_0^3}+4\,{m_x^2\over m^5}\right]dx, \eqno(4.9c)$$
$$\tilde I_3=\int_{-\infty}^\infty\left[{1\over m^5}-{1\over u_0^5}+8\,{m_{xx}^2\over m^7}+20\,{m_x^2\over m^7}-56\,{m_x^4\over m^9}\right]dx. \eqno(4.9d)$$
Note that $\tilde I_0$ follows directly from the mCH equation (1.1).\par
 Last, we  remark that an infinite number of conservation laws
have also  been obtained quite recently using the fact that the mCH equation describes pseudo-spherical surfaces$^{21}$ which is equivalent to the existence of the
Lax representation for the mCH equation.$^5$  On the other hand, our derivation is based on a purely algebraic procedure relying on the B\"acklund transformation. 
\par
\bigskip
\leftline{\bf V. CONCLUDING REMARKS}\par
\bigskip
In this paper, we have developed a systematic procedure for constructing smooth soliton solutions of the mCH equation.
In the process, the associated mCH equation has played a central role which has been derived from the mCH equation  via a reciprocal
transformation. It is important that certain condition must be imposed on the soliton parameters to assure the smoothness of the solutions.
Specifically, a detailed inspection of the one-soliton solution reveals that there exists a critical value of the amplitude parameter beyond which
the solution becomes a many-valued function. This feature differs from that of the one-soliton solution of the CH equation for which the peakon solution is produced from the
smooth soliton solution in the limit of small dispersion.$^{18,19}$  Whether the singular solutions such as the W/M-shape soliton and cusp soliton found by Qiao$^{5}$
can be obtained from smooth solitons by means of appropriate limiting procedure is an interesting issue to be studied in a separate context. \par
An exact method of solution presented here is also applicable to a variant of the mCH equation, for example,
$$ m_t+[m(u^2-u_x^2)]_x+\gamma u_x=0,\quad  m=u-u_{xx}, \eqno(5.1)$$
where $\gamma$ is a real parameter.  When $\gamma=0$, this equation is shown to exhibit peakons under the boundary condition $u\rightarrow 0$ as $|x|\rightarrow \infty$.$^{22}$
In addition, another version of the mCH equation with cubic nonlinearity  
$$m_t+u^2m_x+3uu_xm=0,\quad  m=u-u_{xx}, \eqno(5.2)$$
attracts an interest because of its different mathematical structure from that of the mCH equation under consideration.
  This equation has been discovered by Novikov$^{23}$ and admits a Lax pair related to a negative flow of the Sawada-Kotera hierarchy through
a reciprocal transformation.$^{24}$
For both equations (5.1) and (5
.2), we obtained smooth single soliton solutions which satisfy the boundary condition $u\rightarrow u_0$ as $|x|\rightarrow \infty$.
However, the general $N$-soliton problem still remains open and it reserves a future study.
\par
In conclusion, we comment on a paper by Ivanov and Lyons.$^{8}$ 
Applying the IST to an initial value problem of the mCH equation under the boundary condition $m\rightarrow m_0(=u_0)$ as $|x|\rightarrow \infty$,
they presented the explicit parametric
representations of the
one- and two-soliton solutions for the 
 quantity  $m(=u-u_{xx})$ which take the form of   dark solitons. See Fig. 1 in their paper, for example.
However, the expression (37) for $x$ is incorrect.  Actually, the sign of the first term on the right-hand side of (37) should be minus in place of plus. 
This discrepancy has been caused by an error in Eq. (32). Specifically, a factor ${\rm e}^{iky}$ should read ${\rm e}^{-iky}$. The same error happened in Eq. (37) as well for the
two-soliton solution. If these errors were corrected appropriately, then their result would coincide with that given in the present paper. Note, however that Ivanov and Lyons
obtained the expression of $m$ only. On the other hand, we have a compact parametric representation  of $u$  for the first time. \par
\bigskip
\leftline{\bf ACKNOWLEDGEMENTS}\par
\bigskip
This work was partially supported by JSPS KAKENHI Grant Number 22540228.  \par

\newpage
\leftline{\bf REFERENCES} \par
\baselineskip=6mm
\begin{enumerate}
\item A.S. Fokas, The Korteweg-de Vries equation and beyond, Acta. Appl. Math. {\bf 39}, 295-305 (1995).
\item B. Fuchssteiner, Some tricks from the symmetry-toolbox for nonlinear equations: Generalizations of the Camassa-Holm equation, Phys. D{\bf 95}, 229-243 (1996).
\item P.J. Olver and P. Rosenau, Tri-Hamiltonian duality between solitons and solitary-wave solutions having compact support, Phy. Rev. E{\bf 53}, 1900-1906 (1996).
\item R. Camassa and D.D. Holm, An integrable shallow water equation with peaked solitons, Phy. Rev. Lett. {\bf 71}, 1661-1664 (1993).
\item Z. Qiao, A new integrable equation with cuspons and W/M-shape-peaks solitons, J. Math. Phys. {\bf 47}, 112701 (2006).
\item D.D. Holm and R. Ivanov, Smooth and peaked solitons of the CH equation, J. Phys. A: Math. Theor. {\bf 43}, 434003 (2010).
\item Z. Qiao, New integrable hierarchy, its parametric solutions, cuspons, one-peak solitons, and M/W-shape peak solitons, J. Math. Phys. {\bf 48}, 082701 (2007).
\item R.I. Ivanov and T. Lyons, Dark solitons of the Qiao's hierarchy, J. Math. Phys. {\bf 53}, 123701 (2012).
\item Y. Matsuno, Parametric representation for the multisoliton solution of the Camassa-Holm equation, J. Phys. Soc. Jpn. {\bf 74}, 1983-1987 (2005).
\item Y. Matsuno, Multisoliton solutions of the Degasperis-Procesi equation and their peakon limit, Inverse Prob. {\bf 21}, 1553-1570 (2005).
\item Y. Matsuno, The $N$-soliton solution of the Degasperis-Procesi equation, Inverse Prob. {\bf 21}, 2085-2101 (2005).
\item M.J. Ablowitz, D.J. Kaup, A.C. Newell and H. Segur, The inverse scattering transform - Fourier analysis for nonlinear problems, Stud. Appl. Math. {\bf53}, 249-315 (1974).
\item T.L. Perelman, A.Kh. Fridman and M.M. El'yashevich, On the relationship between the $N$-soliton solution of the modified Korteweg-de Vries equation and the KdV equation solution,
       Phys. Lett. {\bf 47A}, 321-323 (1974).
\item H. Ono, Solitons on a background and a shock wave, J. Phys. Soc. Jpn. {\bf 40}, 1487-1496 (1976).
\item R. Hirota and J. Satsuma, $N$-soliton solutions of model equations for shallow 
water waves, J. Phys. Soc. Jpn. {\bf 40}, 611-612 (1976)
\item R.M. Miura, Korteweg-de Vries equation and generalizations I. A remarkable explicit nonlinear transformation, J. Math. Phys. {\bf 9}, 1202-1204 (1968).
\item  R. Hirota, Direct methods in soliton theory. In {\it Solitons} (ed. R.K. Bullough and P.J. Caudrey, Springer-Verlag, New York, 1980)  157-176.
\item A. Parker and Y. Matsuno, The peakon limits of soliton solutions of the Camassa-Holm equation, J. Phys. Soc. Jpn. {\bf 75}, 124001 (2006).
\item Y. Matsuno, The peakon limit of the $N$-soliton solution of the Camassa-Holm equation, J. Phys. Soc. Jpn. {\bf 76}, 034003 (2007).
\item P.A. Clarkson and E.L. Mansfields, On a shallow water wave equation, Nonlinearity {\bf 7}, 975-1000 (1994).
\item P.M. Bies, P. G\'orka and E.G. Reyes, The dual modified Korteweg-de Vries-Fokas-Qiao equation: Geometry and local analysis, J. Math. Phys. {\bf 53}, 073710 (2012).
\item G. Gui, Y. Liu, P.J. Olver and C. Qu, Wave-breaking and peakons for a modified Camassa-Holm equation, Commun. Math. Phys. (2012), {\bf 319}, 731-759 (2013).
\item V. Novikov, Generalizations of the Camassa-Holm equation, J. Phys. A: Math. Theor. {\bf 42}, 342002 (2009).
\item A.N.W. Hone and J.P. Wang, Integrable peakon equations with cubic nonlinearity, J. Phys. A: Math. Theor. {\bf 41}, 372002 (2008).

\end{enumerate}

\end{document}